\documentclass[runningheads]{llncs}
\usepackage{todonotes}
\usepackage{xcolor}
\usepackage{mathtools}
\usepackage{multirow}
\usepackage{algorithm}
\usepackage{algorithmic}
\usepackage{adjustbox}
\usepackage{graphicx}
\usepackage{todonotes}
\usepackage{xcolor}
\usepackage{mathtools}
\usepackage{multirow}
\usepackage{algorithm}
\usepackage{algorithmic}
\usepackage{adjustbox}
\usepackage{graphicx}
\usepackage{amsfonts}
\usepackage{enumitem}

\usepackage{amsfonts}
\begin{document}
\title{Leveraging Application-Specific Knowledge for Energy-Efficient Deep Learning Accelerators on Resource-Constrained FPGAs}
\titlerunning{Application-Aware Energy-Efficient Deep Learning Accelerators on FPGAs}
\author{Chao Qian\orcidID{0000-0003-1706-2008}}
\authorrunning{C. Qian}
\institute{Intelligent Embedded Systems Lab,\\ 
University of Duisburg-Essen, Germany \\
\email{chao.qian@uni-due.de}}
\maketitle             
\begin{abstract}
The growing adoption of Deep Learning (DL) applications in the Internet of Things has increased the demand for energy-efficient accelerators. Field Programmable Gate Arrays (FPGAs) offer a promising platform for such acceleration due to their flexibility and power efficiency. However, deploying DL models on resource-constrained FPGAs remains challenging because of limited resources, workload variability, and the need for energy-efficient operation.

This paper presents a framework for generating energy-efficient DL accelerators on resource-constrained FPGAs. The framework systematically explores design configurations to enhance energy efficiency while meeting requirements for resource utilization and inference performance in diverse application scenarios.

The contributions of this work include: 
(1) analyzing challenges in achieving energy efficiency on resource-constrained FPGAs; (2) proposing a methodology for designing DL accelerators with integrated Register Transfer Level (RTL) optimizations, workload-aware strategies, and application-specific knowledge; and (3) conducting a literature review to identify gaps and demonstrate the necessity of this work.

\keywords{FPGA \and Deep Learning \and Energy-Efficient \and Accelerator}
\end{abstract}

\section{Introduction}
\label{sec:introduction}

The rapid growth of \emph{Deep Learning} (DL) in the \emph{Internet of Things} (IoT) has revolutionized domains such as smart homes, healthcare, and industrial automation~\cite{cheng2024advancements}. By enabling IoT devices to process complex data and make intelligent decisions, DL has unlocked new possibilities for autonomous and real-time operations. However, these advancements are constrained by the physical size, power, and energy limitations of IoT devices, which challenge the deployment of computationally intensive DL models on \emph{Microcontrollers} (MCUs). This creates a critical need for compact, energy-efficient hardware accelerators that balance computational performance with these constraints.

\emph{Field Programmable Gate Arrays} (FPGAs) have emerged as a promising solution for deploying DL models on embedded platforms. They offer high flexibility for hardware customization and significant power efficiency, making them suitable for resource-constrained IoT devices. However, deploying DL models on FPGAs presents several challenges. Limited on-chip resources and high memory demands must be addressed while maintaining energy efficiency and performance. Additionally, selecting an appropriate FPGA size involves trade-offs: larger FPGAs consume more static power, while smaller FPGAs may lack the capacity to accommodate complex models. Frequent reconfiguration in duty-cycled operation modes, where the FPGA is turned off when not needed, introduces additional inefficiencies and makes energy-efficient DL inference even more difficult.

My PhD research focuses on the following questions to improving the energy efficiency of DL accelerators on FPGAs:\\
\textbf{(RQ1)} How can hardware accelerators be designed at the Register Transfer Level (RTL) to effectively utilize model-level optimizations, such as selecting suitable activation function implementations, to achieve energy-efficient inference on FPGAs? \\
\textbf{(RQ2)} What workload-aware strategies can be implemented to adapt inference efficiency dynamically to various workload demands? \\
\textbf{(RQ3)} How can application-specific knowledge be utilized to combine RTL optimizations and workload-aware strategies to derive the most energy-efficient DL accelerator?

Guided by the above questions, this paper proposes a systematic methodology for designing energy-efficient, problem-specific DL accelerators tailored to resource-constrained FPGAs. The approach integrates optimized RTL templates, workload-adapted execution strategies, and application-specific knowledge within a flexible framework. This methodology aims to maximize system energy efficiency while meeting the constraints defined by the application.

The remainder of this paper is structured as follows: Section~\ref{sec:research_methodology} presents the proposed methodology, detailing the steps used to address the research questions. Section~\ref{sec:current_state_of_research} discusses my current progress and findings. Section~\ref{sec:future_work} outlines the planned work for completing my PhD. Section~\ref{sec:related_work} positions my work within the context of related research. Finally, Section~\ref{sec:conclusion} summarizes the key findings and contributions.

\section{Research Methodology}
\label{sec:research_methodology}
\begin{figure}[!htb]
    \centering
    \includegraphics[width=1\linewidth]{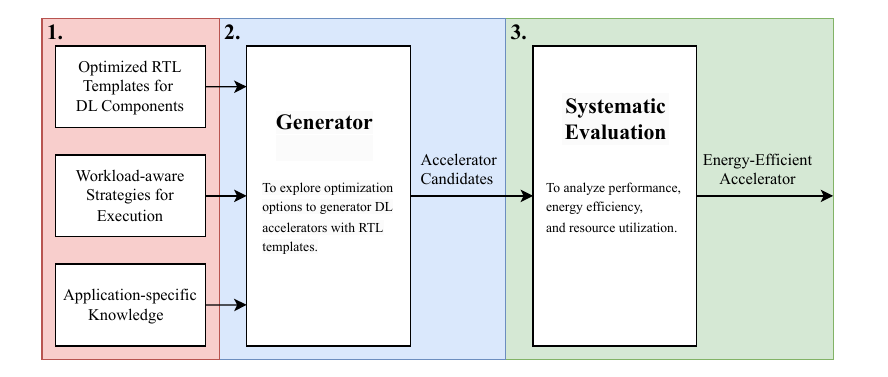}
    \caption{Conceptual framework illustrating how RTL templates, workload-aware strategies, and application requirements are utilized to generate and evaluate energy-efficient DL accelerator designs.}
    \label{fig:conceptual_framework}
\end{figure}

This section outlines my research methodology to design energy-efficient DL accelerators for FPGAs, guided by the conceptual framework depicted in Figure~\ref{fig:conceptual_framework}. The methodology integrates three key steps: (1) preparing optimized RTL templates, workload-aware strategies, and application-specific knowledge as inputs; (2) combining these inputs within a \emph{Generator} to produce optimized accelerator candidates; and (3) evaluating the candidates to identify the most efficient design.

\subsection{Inputs for Accelerator Generation}
\label{subsec:inputs}

The framework begins with three key inputs, as illustrated in Figure~\ref{fig:conceptual_framework}:

\begin{itemize}
    \item \textbf{Optimized RTL Templates for DL Components:} 
    These templates provide reusable hardware designs for core DL operations, such as activation functions, fully connected layers, and \emph{Long Short-Term Memory} (LSTM) layers. Each operation can be implemented in multiple ways to achieve different optimization goals, such as minimizing resource usage or maximizing clock frequency. Selecting the most suitable implementation depends on application-specific needs and resource availability. 

    \item \textbf{Workload-aware Strategies:} 
    Workload-aware strategies manage the unique characteristics of IoT devices, where sensor data collection is often slower than FPGA inference. Strategies include powering off FPGA with reconfiguration overhead or keeping it active to avoid reconfiguration overhead, which can be used to minimize energy consumption during idle periods. Alternatively, the inference speed can be reduced to align the inference time with the request period, preventing idle states and reconfiguration inefficiencies.

    \item \textbf{Application-Specific Knowledge:} 
    Application-specific knowledge defines optimization goals (e.g., maximizing energy efficiency) and constraints (e.g., latency thresholds or resource limits). By aligning optimization goals with constraints, the framework ensures accelerators are efficient and practical for deployment.
\end{itemize}

These inputs lay the groundwork for the \emph{Generator}, which will be detailed in the following sections.

\subsection{Generator}
\label{subsec:generator}

The \emph{Generator} produces optimized DL accelerators by systematically exploring the design space defined by the inputs in Section~\ref{subsec:inputs}. Its process includes:

\begin{itemize}
    \item \textbf{Defining the Design Space:} The \emph{Generator} uses performance profiles from optimized RTL templates and workload adaptation strategies to establish exploration boundaries.

    \item \textbf{Exploration and Estimation:} The \emph{Generator} prioritizes one metric, such as energy efficiency, as the optimization goal while treating others, like latency and resource utilization, as constraints. Analytical models estimate the performance of candidate accelerators, allowing early pruning of suboptimal designs.

    \item \textbf{Generating Outputs:} Multiple accelerator candidates are produced, each representing a unique configuration that meets the defined constraints. These candidates are evaluated in the next phase to identify the most suitable design.
\end{itemize}

This structured exploration ensures an efficient traversal of the design space, focusing on configurations most likely to meet energy efficiency and performance requirements.

\subsection{Systematic Evaluation}
\label{subsec:systematic_evaluation}

The evaluation phase validates the impacts of individual inputs and their combination on the final accelerators. It employs the following tools and methods:

\begin{itemize}
    \item \textbf{Evaluation Tools:}
        \begin{itemize}
            \item \textbf{Behavior Simulation:} Tools like GHDL verify the mathematical correctness and functionality of accelerators and calculate inference time in clock cycles.
            \item \textbf{Electronic Design Automation (EDA) Tool Analysis:} FPGA vendor tools, such as AMD Vivado and Lattice Radiant, generate reports on resource utilization, power consumption, and timing performance. Besides, these estimations can be conveniently replicated by other researchers, thanks to the widespread availability and adoption of these tools.
            \item \textbf{Real Hardware Measurements:} Hardware platforms measure energy consumption, throughput, and latency under practical conditions.
        \end{itemize}

    \item \textbf{Progressive Evaluation:}
        \begin{itemize}
            \item \textbf{Standalone Input Evaluation:} Each input, such as RTL templates or workload-aware strategies, is evaluated independently to isolate its contribution to energy efficiency and performance, answering RQ1 and RQ2.
            \item \textbf{Combined Optimization Evaluation:} Accelerators generated using all inputs are evaluated to address RQ3, verifying whether their combination results in superior energy efficiency and performance.
        \end{itemize}
\end{itemize}

Including simulation with EDA tools and testing on real hardware offers a chance for cross-checking. Furthermore, the progressive evaluation approach allows adjustments to inputs and the \emph{Generator}, enabling refinement of the design process and minimizing the risk of unresolved research questions.

\section{Current State of Research}
\label{sec:current_state_of_research}

This section outlines my progress toward developing energy-efficient DL accelerators for FPGAs. Substantial advancements have been achieved in three key areas: optimized RTL templates for DL components, workload-aware strategies, and evaluation infrastructure.

\subsection{Optimized RTL Templates}
To address RQ1, I have made progress in developing optimized RTL templates for core DL operations, including LSTM cells, Convolutional layers, fully connected layers, and attention modules in Transformer models. These templates are designed to improve energy efficiency while ensuring high performance.

For the LSTM accelerator~\cite{qian2024exploring}, notable improvements were achieved in both latency and energy efficiency through pipelining and activation function optimization at the RTL level. Latency was reduced from 53.32 $\mu$s to 28.07 $\mu$s, representing a 47.37\% reduction. Energy efficiency improved from 5.57 GOPS/s/W to 12.98 GOPS/s/W, marking a 2.33$\times$ increase.

Similarly, accelerators for \emph{Convolutional Neural Networks}~\cite{burger2020embedded} and \emph{Multilayer Perceptrons} (MLPs)~\cite{ling2024configurable} with template optimizations have been validated through analytical models and hardware tests. These results further demonstrate the capability of optimized RTL templates to improve performance and energy efficiency, meeting the stringent constraints of resource-constrained FPGAs.

Additionally, activation functions such as Sigmoid, Tanh, HardSigmoid, and HardTanh have been optimized to provide multiple implementation options~\cite{qian2024exploring,qian2022enhancing}. These variations enable trade-offs between precision, resource usage, and throughput, allowing designers to select the most suitable implementation for specific application requirements.

\subsection{Workload-Aware Optimization}
To address RQ2 of my research, I have focused on workload-aware optimization, which tackles runtime inefficiencies by adapting accelerators to varying workload conditions. By optimizing the FPGA configuration phase and implementing the Idle-waiting strategy~\cite{qian2024idle}, substantial energy savings and improved workload management have been achieved.

For regular request periods, the Idle-Waiting strategy demonstrated superior energy efficiency compared to the traditional On-Off approach~\cite{qian2024idle}. During a 40 ms request period, this strategy processed 12.39$\times$ more workload items within the same energy budget, effectively extending the system lifetime and addressing challenges posed by shorter request intervals.

To address irregular workloads, I have developed an adaptive strategy-switching mechanism using predefined and learnable thresholds~\cite{qian2025configuration}. The learnable threshold method outperformed the predefined approach with a 6\% performance improvement, providing a robust and efficient solution for dynamic workload management.

These advancements indicate the feasibility of utilizing workload-awareness to improve the system energy efficiency for  FPGA-based platforms.

\subsection{Evaluation Infrastructure}
The evaluation of accelerators begins with software-based analysis using tools such as AMD Vivado and Lattice Radiant. EDA tools provide reports with insights into resource utilization, power estimations, and timing performance.

Based on these software-based insights, the \textit{Elastic Node} platform has been iterated within our research group over the past five years as a dedicated hardware testbed~\cite{burger2020elastic,qian2023elasticai}. This platform is used for real-world validation. It measures metrics such as energy consumption, throughput, and latency under practical conditions, further validating the reports from EDA tools.

By combining software insights with hardware measurements~\cite{qian2023elasticai}, this approach ensures realistic evaluations of accelerator designs and enables accessible performance comparisons.

\section{Future Work}
\label{sec:future_work}

Future work will address the remaining challenges to fully validate the research questions and further enhance the methodology for designing energy-efficient DL accelerators on FPGAs. Key focus areas include integrating and identifying inputs, implementing search algorithms, and rigorously evaluating the proposed framework.

The next step will fully integrate optimized RTL templates, workload-aware strategies, and application-specific knowledge into the \emph{Generator} framework. Prioritizing inputs based on their impact on energy efficiency and developing adaptive mechanisms for dynamic inclusion will ensure that the \emph{Generator} remains flexible and adaptable to varying requirements.

In parallel, I will implement search algorithms to explore combinations of inputs, such as RTL templates and workload strategies, while considering application-specific constraints. Finally, thorough evaluations will quantify the impact of application-specific knowledge on energy efficiency improvement. 

This process includes assessing the individual and combined contributions of inputs to overall system performance and demonstrating energy efficiency improvements by comparing the designs generated by my methodology against baseline implementations under diverse workload conditions.

\section{Related Work}
\label{sec:related_work}

Developing energy-efficient DL accelerators for resource-constrained FPGAs involves three key research areas: hardware optimization, workload adaptation, and search algorithms. This section reviews contributions in these areas and positions this work within the broader context of energy-efficient acceleration.

\subsection{Hardware Optimization for Deep Learning Components}
\label{hardware_optimization_for_dl_components}
My PhD research builds on an earlier study in resource reuse techniques aimed at developing energy-efficient accelerators. Schiele et al.~\cite{schiele2019elastic} introduced an MLP accelerator implemented on Spartan-6 LX9 FPGAs, delivering significant energy efficiency improvements over low-power MCUs. This design supported model training on the FPGA but was limited to an operating frequency of 50 MHz due to the backward propagation complex of the design. Subsequent efforts simplified the design by removing backward propagation, limiting it to feedforward propagation. Utilizing the newer Spartan-7 XC7S15 FPGA, the updated MLP accelerator achieved a clock frequency of 100 MHz for a soft sensor application~\cite{ling2023device}.

Research on LSTM accelerators has evolved significantly, with distinct approaches to arithmetic unit allocation. Some studies focused on parallelizing all \emph{Arithmetic Logic Units} (ALUs), maximizing throughput but resulting in inefficient resource utilization~\cite{cao2019efficient,rybalkin2018finn}. Conversely, other works prioritized resource efficiency by implementing minimal ALUs and reusing them over time~\cite{manjunath2020low,chen2021eciton}. While the latter showcased superior resource efficiency, its energy efficiency suffered due to prolonged execution times.

Within neural networks, particularly LSTMs, activation functions play a crucial role. Early studies~\cite{li2022fpga,pan2022modular,pogiri2022design,shatravin2022sigmoid} explored implementing functions like Sigmoid and Tanh on FPGAs, emphasizing resource efficiency and precision. As quantization-aware training gained traction, recent works demonstrated the viability of simplified activation functions, such as HardSigmoid and HardTanh, which achieve no precision loss between software definitions and hardware implementations while significantly reducing computational overhead~\cite{manjunath2020low,qian2023energy}.

The impact of precision on energy efficiency has also been studied as a key factor in FPGA-based accelerator optimization. Rybalkin et al.~\cite{rybalkin2018finn} systematically explored the design space concerning precision for Bidirectional Long Short-Term Memory (BiLSTM) neural networks. Their study highlights that significantly reducing precision enhances hardware efficiency, improving memory usage, energy consumption, and throughput. However, to our knowledge, no existing work has systematically prioritized energy efficiency as the primary optimization goal while exploring the design space for accelerators.

\subsection{Workload-Aware Optimization}

Studies in Section~\ref{hardware_optimization_for_dl_components} focus on optimizing inference phases for continuous processing tasks where the FPGA remains busy. However, in practical IoT applications, DL tasks often involve discontinuous workloads, resulting in idle periods. One method is to power off the FPGA during these periods to avoid idle power consumption. However, it introduces configuration overhead because the FPGA must be reconfigured each time it powers back on, which adds time and energy costs, potentially offsetting the overall energy efficiency.

Some researchers have explored optimizing the FPGA configuration process to address this challenge. Fritzsch et al.~\cite{fritzsch2022reduction} proposed compressing the bitstream by 1.05× to 12.2× to reduce configuration time, but they did not evaluate its impact on energy efficiency. Similarly, Cichiwskyj et al.\cite{cichiwskyj2020time} introduced Temporal Accelerators, showing that even when the accelerator is split into two bitstreams, requiring to configuring the FPGA two times, a smaller FPGA (Spartan-7 XC7S6) could achieve greater energy efficiency than a larger one (Spartan-7 XC7S15) for a single inference.
However, these studies have not utilized the workload intensity to change the configuration behavior, which can be applied to improve the system's energy efficiency.

\subsection{Research Gap and Positioning}

Despite progress in hardware optimization and workload-aware strategies, key gaps persist. Existing methods often lack integration of application requirements into the accelerator design process and fail to effectively address dynamic workload adaptation in heterogeneous platforms combining MCUs and FPGAs. Additionally, efficient exploration of optimal configurations under constraints such as resource constraints and workload variability remains underexplored.

My PhD aims to address these challenges through a systematic methodology, utilizing a \emph{Generator} that leverages application-specific knowledge to guide design space exploration and maximize the energy efficiency of DL accelerators.


\section{Conclusion}
\label{sec:conclusion}

This research hypothesizes incorporating application-specific knowledge into a generator framework can produce DL accelerators with enhanced energy efficiency. The achievements to date include the development of efficient hardware templates for DL components, implementing workload-aware strategies, and establishing a robust evaluation platform. Preliminary progress has been demonstrated in enhancing energy efficiency for LSTM accelerators and developing strategies to manage both regular and irregular workloads effectively.

This study has also identified key gaps in existing research, including the limited integration of application-specific knowledge and the lack of systematic exploration algorithms for optimal accelerator design. 

The next steps will focus on completing the proposed methodology and validating the feasibility of applying application-specific knowledge to derive optimal accelerator configurations automatically.

\vspace{24 pt}
\noindent{\textbf{Acknowledgement.}} The author gratefully acknowledges the supervision Prof. Dr. Gregor Schiele.
\vspace{24 pt}
\bibliographystyle{IEEEtran}
\bibliography{references}
\end{document}